\documentclass[9pt, a4paper, final, twoside]{extarticle}  

\usepackage[headings,cm]{fullpage}

\usepackage{times,url,pst-node,parallel,amssymb,mdwlist,marvosym,fancyhdr}

\usepackage[afrikaans,british,hungarian]{babel}

\usepackage[OT2,OT1]{fontenc}
\def\cyr{\fontencoding{OT2}\fontfamily{wncyr}\selectfont}

\newcommand{\C}{{\mathbb C}}

\title{\flushleft \vskip 1cm 
\Huge {\bf Die voorgeskiedenis van kwantumberekening}\\
\large \bigskip
\rm \textsf{P.H. Potgieter}\thanks{
Die outeur wil die stigting Pro Renovanda Cultura Hungariae Alap\'itv\'any
vir sy ondersteuning en die Departement Wiskunde van die Boedapestse Universiteit vir Ekonomiese 
Wetenskappe en Publieke Administrasie vir sy ondersteuning en gasvryheid bedank wat 
'n besoek aan laasgenoemde, waartydens 'n groot deel van die inligting hier vervat ingewin is, 
in die laaste deel van 2003 moontlik gemaak het.}
\\ 
\rm Departement Kwantitatiewe Bestuur, Unisa, Posbus 392, Unisarand 0003 \\
\rm E-pos: {\tt potgiph@unisa.ac.za}}

\date{}

\begin{document}

\selectlanguage{afrikaans}

\hyphenation{ander-sins kwantum-ou-tomaat Cyber-netics teorie}

\maketitle
\thispagestyle{fancy}

\noindent
{\bf \Large \textsc{Uittreksel}}\\
\emph{
Die hoofidees wat tans gestalte vind in die teorie en tegnologie van kwantumberekening is 
in die laat 1970's en vroeg 1980's deur fisici in die Weste en 
'n wiskundige in die voormalige Sowjetunie neergel\^e. Dat di\'e teorie ook wortels in die 
Russiestalige vakliteratuur het, is nie algemeen bekend in die Weste nie. 
Daar word kortliks gekyk na die idee soos deur Benioff en (veral) Feynman in die Weste 
versprei, asook die voorstel van di\'e rekengrondslag deur Manin in die 
Russiese literatuur. 
Die outeur hoop om hiermee so 'n onpartydig moontlike sintese van die 
vroe\"e gedagtegeskiedenis rondom 
kwantumberekening aan te bied. Die rol van omkeerbare en onomkeerbare berekeningsprosesse word 
vlugtig bekyk soos dit verband hou met die ontstaan van kwantumberekening, asook die sogenaamde 
Inligtingsparadoks in die fisika. Die inligtingsteorie en die fisika het heelwat met 
mekaar te kommunikeer, soos hierdie paradoks uitwys.}

\bigskip

\selectlanguage{british}\noindent
{\Large\bf\textsc{Abstract}}\\
{\bf\emph{The pre-history of quantum computation}}\\
\emph{The main ideas behind developments in the theory and technology of quantum computation 
were formulated in the late 1970s and early 1980s by two physicists in the West and 
a mathematician in the former Soviet Union. It is not generally known in the West 
that the subject has roots in the Russian technical literature. The idea, as propagated 
by Benioff and (especially) Feynman, is reviewed along with the proposition of a 
foundation for this kind of computation by Manin in the Russian literature. The author 
hopes to present as impartial a synthesis as possible of the early history of thought 
on this subject. The role of reversible and irreversible computational processes will 
be examined briefly as it relates to the origins of quantum computing and the so-called 
Information Paradox in physics. Information theory and physics, as this paradox shows, 
have much to communicate to each other.}

\setlength{\parindent}{12pt}

\selectlanguage{afrikaans}

\section{\textsc{Inleiding}}

Kwantumberekening is 'n betreklik nuwe dissipline wat baie belofte inhou vir 'n vinnige 
en doeltreffende\,---\,in terme van ruimte \'en energie\,---\,toevoeging tot ons arsenaal van 
rekenapparatuur. Daar is in die laaste tien jaar algoritmes ontwikkel vir priemfaktorisering 
in polinoomtyd deur Peter Shor by Bell Labs (1994) en vir 'n soektog deur 'n ongeordende lys deur 
Lev Grover\,---\,ook van Bell Labs\,---\, in 1996 \cite{Grover1996}. 'n Goeie oorsig oor die stand van 
kwantumalgoritmes is Peter Shor se aantekeninge \cite{Shor2000}. 
Die tegnologie van kwantumrekenmasjiene is in sy 
kinderskoene, met min praktiese suksesse soos byvoorbeeld die priemfaktorisering in 2001  deur IBM 
van die getal 15 met behulp van 'n kwantumrekenaar. Fisici verwag egter dat daar 
oor die volgende paar jaar duidelike vordering sal wees en ondersoekspanne in onder andere  
die VSA, Duitsland, Japan en Australi\"e werk aan 'n aantal verskillende benaderings om 
di\'e toestelle te realiseer. Weens die ho\"e nut van 'n vinnige, praktiese 
priemfaktoriseringsalgoritme om die openbare sleutelkriptografie, soos 
algemeen gebruik word deur banke en deur individue, te breek, behoort in gedagte 
gehou te word dat vordering op hierdie gebied nie noodwendig sonder 
geheimhouding sal gebeur nie. 
In hierdie bydrae sal gekyk word na die idees wat gelei het tot die formulering van 'n model 
vir kwantumberekening, veral na die afsonderlike en feitlik gelyktydige verskyning van 
di\'e idees in die VSA en in die Sowjetunie.

\section{\textsc{Formele begrippe van berekenbaarheid}}

Die sterk behoefte om 'n formele definisie van \emph{algoritme} of \emph{berekening} te gee, het aan 
die wetenskaplike gemeenskap aan die begin van die twintigste eeu duidelik geword hoofsaaklik 
as gevolg van twee probleme wat deur Hilbert gestel is:
\begin{list}{\textbullet}{\rightmargin 0,75cm}
\item die \emph{Entscheidungsproblem} of Beslissingsprobleem: Bestaan daar 'n algoritme s\'o 
dat dit, gegee 'n willekeurige stelling in eerste-orde-logika (byvoorbeeld Peano-rekenkunde), 
kan bepaal of di\'e stelling waar is in alle modelle van 'n teorie, al dan nie? en
\item Hilbert se Tiende Probleem: Bestaan daar 'n algoritme wat, gegee 'n Diofantiese 
vergelyking, kan bepaal of die vergelyking enige heeltallige oplossings het, al dan nie?
\end{list}
Die Beslissingsprobleem kan teruggevoer word na Leibniz en is onafhanklik suksesvol opgelos in die 
middel 1930's deur Alonzo Church en Alan Turing. Church het enige funksie wat deur sy 
Lambda-rekene gedefinieer kon word as \emph{algoritme} gedefinieer. Turing het 
eers die begrip \emph{algoritme} as identies met \emph{rekursiewe funksie} geneem  
en later as funksie wat bereken word deur 'n sogenaamde Turing-masjien\footnote{
Die nuuskierige leser sal nabootsers vir Turing-masjiene op die Internet vind, o.a. 
op \url{http://www.nmia.com/~soki/turing/}.}\,---\,'n ge\"\i dealiseerde toestel wat vandag nog 
voorgraadse studente laat kopkrap. Dit is later aangetoon dat die klasse van 
die rekursiewe funksie, funksies van die Lambda-rekene en die Turing-masjienberekenbare 
funksies een en dieselfde is. Hierdie\,---\,toe ietwat verbasende\,---\,gelykstelling van drie 
definisies met heel verskillende herkoms het die gedagte dat d\'it 'n toereikende 
definisie van berekenbaarheid is, sterk ondersteun. Die Beslissingsprobleem word dan 
ook, in die lig van die bevredigende definisie van 'n algoritme, as opgelos beskou.

\section{\textsc{Omkeerbare en onomkeerbare berekening}}

Aanvanklik is berekening as 'n essensieel onomkeerbare proses (minstens in die praktyk)  
beskou. 
'n Berekening word \emph{omkeerbaar} genoem indien daar 'n een-tot-een-verwantskap tussen 
die versameling van alle moontlike uitvoere en di\'e van alle moontlike invoere van die 
proses bestaan. Dit impliseer dat die invoer van 'n omkeerbare rekenproses altyd uit die uitvoer 
herwin kan word. Rekenprosesse in die alledaagse lewe is nie omkeerbaar nie. Ons voer 
byvoorbeeld 'n heelgetal in en ontvang as uitvoer {\bf 1} indien die invoer 'n priemgetal 
is en {\bf 0} andersins\,---\,die invoer kan beslis hier nie uit die uitvoer bereken word nie. 
Eintlik is wat hier beskryf word \emph{logiese} omkeerbaarheid, in teenstelling met \emph{fisiese}  
omkeerbaarheid wat gekarakteriseer word daardeur dat die entropie nie toeneem nie. 
Rolf Landauer het egter in 1960 al tot die gevolgtrekking gekom dat 
elke logies omkeerbare berekening ook in beginsel fisies omkeerbaar verwerklik kan word, 
dit wil s\^e sonder 'n toename in entropie \cite{VLi}. Landauer het die minimale toename in entropie 
wat volg uit die vernietiging van een bis se inligting ge\"\i dentifiseer \cite{Lloyd1999}. Die 
toename in entropie is minstens $(\ln 2) k$ waar $k$ Boltzmann se konstante is:  
$k\approx 1,38 \times 10^{-23}JK^{-1}$. 
Die energie wat vereis word om een bis uit te wis teen 'n konstante temperatuur $T$ 
is dus rofweg $kT \ln 2$. 
Charles Bennett \cite{Bennett1982} het Landauer se termodinamiese analise 
gebruik om die skynparadoks van Maxwell se Duiweltjie op te los. Bennett se oplossing 
is strydig met di\'e van Szillard (in 1929), omdat sy argument juis daarop berus dat di\'e 
Duiweltjie inligting \emph{vernietig}.

\subsection{Die Inligtingsparadoks}

Die \emph{Inligtingsparadoks} vir swart gate handel oor die inligtingsverlies wat moontlik 
gepaardgaan met die inval van 'n voorwerp of stelsel, s\^e maar die goue plaat op die 
ruimteskip Voyager 2, in 'n swart gat. Volgens die teoretiese astrofisika kan 'n swart 
gat geen ander eienskappe as massa, hoekmomentum en elektriese lading vertoon nie en 
derhalwe sou die inligting op die goue plaat o\"enskynlik verdwyn het. 
Volgens Stephen Hawking, wat baanbrekerswerk oor swart gate gedoen het, gaan di\'e inligting 
eenvoudig verlore. (Sien byvoorbeeld \cite{Hooft1995} vir 'n meer volledige bespreking.) 
Indien d\'it wel is wat gebeur, dan is die styging in temperatuur wat deur Landauer se werk 
 ge\"\i mpliseer word, nie voorhande nie en daar word dus energie vernietig in die swart gat (se 
omgewing) wat strydig is met die wet van energiebehoud. Hawking s\^e:\footnote{
\url{http://www.hawking.org.uk/text/public/dice.html}}
\begin{quote}
{What all this means is, that information will be lost from our region of the universe, 
when black holes are formed, and then evaporate. This loss of information will mean that we 
can predict even less than we thought, on the basis of quantum theory.}
\end{quote}
Hierdie uitgangspunt impliseer dat swart gate nie deur die kwantummeganika (waar alle 
prosesse \emph{omkeerbaar} is) beskryf word nie. 
Die volgende moontlikheid, soos voorgestaan deur Gerard 't Hooft (Nobel-pryswenner 1999, 
Universiteit Utrecht), is dat die inligting wel die swart gat verlaat. D\'it kan 
gebeur \'of deur middel van die Hawking-straling wat uiteindelik, volgens Hawking se 
teorie, die swart gat self tot niet laat gaan, \'of deurdat die inligting op die 
gebeurtenishorison op 'n manier vasgevang word deur snare (\emph{strings}) in 
dimensies anders as die gebruiklike ruimte-tyd. Die laasgenoemde oplossing is een van 'n aantal 
nuwe fisiese teorie\"e wat die kwantummeganika nuut sou kon begrond en voorts 
op nie-omkeerbare basis.

\subsection{Omkeerbare rekenskemas}

John von Neumann het in die 1950's gespekuleer dat elke \emph{logiese} 
bewerking wat deur 'n rekenaar uitgevoer word teen temperatuur $T$ noodwendig minstens 
energie van $kT \ln 2$ moet verstrooi \cite{Lloyd2000}. D\'it is nie strydig met 
Landauer se latere bevinding nie, maar is glad nie in die algemeen waar nie. In 1973 
kon Charles Bennett bewys dat alle Turing-masjienberekeninge 
gedoen kan word deur \emph{omkeerbare} Turing-masjiene en dat geen 
verstrooiing van energie dus noodwendig is nie \cite{Bennett1973}. Bennett het gebruik 
gemaak van Lecerf-omkering, soos ingevoer in 1963 deur Yves Lecerf \cite{Lecerf}, in 
sy beskrywing van 'n tipe omkeerbare Turing-masjien. Saam met die resultaat van Landauer 
beteken d\'it dat dit nie berekening self is nie, maar die uitvee of verlies aan 
inligting wat die entropie laat styg en hitte genereer.

Dit is onafhanklik deur Tommaso Toffoli en Edward Fredkin aangetoon dat elke logiese stroombaan 
vervang kan word deur 'n omkeerbare stroombaan \'en voorts dat slegs een 
omkeerbare logiese hek voldoende is\,---\,die \emph{Toffoli-hek}. Di\'e hek bereken 
$(x,y,z) \mapsto (x,y,z \oplus xy)$ waar $\oplus$ optelling modulo 2 is. Dit staan ook 
bekend as 'n beheerde-beheerde-NIE-hek (\emph{controlled-controlled NOT gate}, figuur 1) waar 
$x$ en $y$ die beheerelemente is.
\begin{figure}[h]
\[\begin{pspicture}(0,-1)(2,1.5) 
\rput(2,-0.75){\pnode{1}{}}
\rput(2,0.5){\pnode{2}{}}
\rput(2,1.75){\pnode{3}{}}
\rput(1,-0.75){\pnode{4}{}}
\rput(1,0.5){\pnode{5}{}}
\rput(1,1.75){\pnode{6}{}}

\rput(0.5,-0.75){\pnode{7}{}}
\rput(0.5,0.5){\pnode{8}{}}
\rput(0.5,1.75){\pnode{9}{}}
\rput(-0.75,-0.75){\pnode{a}{}}
\rput(-0.75,0.5){\pnode{b}{}}
\rput(-0.75,1.75){\pnode{c}{}}

\rput(0.5,-1){\pnode{i}{}}
\rput(1,-1){\pnode{ii}{}}
\rput(0.5,2){\pnode{iii}{}}
\rput(1,2){\pnode{iv}{}}

\ncline[linecolor=black]{i}{ii} \tbput{}
\ncline[linecolor=black]{iii}{iv} \tbput{}
\ncline[linecolor=black]{ii}{iv} \tbput{}
\ncline[linecolor=black]{iii}{i} \tbput{}

\ncline[linecolor=black,arrows=->]{1}{4} \tbput{$z$} 
\ncline[linecolor=black,arrows=->]{2}{5} \tbput{$y$} 
\ncline[linecolor=black,arrows=->]{3}{6} \tbput{$x$}

\ncline[linecolor=black,arrows=->]{7}{a} \tbput{$z\oplus xy$} 
\ncline[linecolor=black,arrows=->]{8}{b} \tbput{$y$} 
\ncline[linecolor=black,arrows=->]{9}{c} \tbput{$x$} 

\end{pspicture}\]
\caption{Die Toffoli-hek}
\end{figure}

\noindent
In die kwantummeganika is 'n stelsel onderhewig aan 'n evolusie wat volledig  
tydomkeerbaar is tot op di\'e stadium dat 'n meting gemaak word. Derhalwe sou 'n 
kwantumstelsel slegs (maar wel!) 'n omkeerbare berekeningsproses kon uitvoer.  
Die ontdekkings van Bennett, Lecerf en Landauer het die skeiding tussen die kwantummeganika 
en die teorie van berekening oorbrug.

\section{\textsc{Feynman, Benioff en Deutsch}}

Richard Feynman het, met sy voorspraak vir die ontwikkeling van 'n teorie 
en tegnologie vir kwantumberekening, onteenseglik 'n reuse-invloed op die 
wetenskaplike publiek gehad. Vanaf 1982 (in \cite{Feynman1982} en in openbare 
lesings o.a.) wys hy nie net uit dat die steeds voortdurende 
verdwerging van elektroniese komponente ons\,---\,vroe\"er of later\,---\,met kwantumeffekte 
in ons rekentoestelle sal 
konfronteer nie, maar ook dat 'n kwantumstelsel 'n baie meer kompakte voorstelling 
van inligting toelaat as 'n konvensionele bis-skikking. Selfs 'n kwantumstelsel 
met net twee \emph{waarneembare} toestande het natuurlik oneindig veel toestande 
in die kwantumtoestandruimte. Voeg ons nog 'n tweetoestand-kwantumstelsel by, dan 
word die nuwe saamgestelde stelsel (in teenstelling met 'n klassieke stelsel) 
beskryf deur 'n toestand in 'n tensorproduk-ruimte, wat die welbekende 
interferensie- en superposisieverskynsels toelaat. Dit is juis hier waar die krag 
van die kwantumberekening sal l\^e. Feynman was in 'n groot mate gemotiveer deur die 
feit dat kwantumstelsels nie sonder eksponensi\"ele vertraging op konvensionele 
rekenmasjiene nageboots kan word nie. Kwantumrekenaars sou dus nodig wees vir die 
doeltreffende simulasie van kwantumstelsels.
 
Die werk van Paul Benioff in 1980--1982 is heelwat minder bekend as Feynman se 
popularisasie. In sy artikel in 1980 \cite{Benioff1980} (en in 1982 in \cite{Benioff1982}) 
beskryf Benioff vir die eerste keer 'n 
kwantummeganiese model vir 'n (omkeerbare) Turing-masjien. Die uitdruklike doel is nie om 
'n nuwe rekenparadigma daar te stel nie, maar om 'n beskrywing van 'n rekenmasjien te 
gee in die terme van die mees fundamentele bekende fisika. Hy laat egter die moontlikheid 
oop dat die kwantumbeskrywing van rekenmasjiene wel gevolge sal h\^e vir die 
werking van di\'e toestelle. 
In 1985 het David Deutsch 'n \emph{universele} kwantumrekenmasjien beskryf \cite{Deutsch1985} 
vir di\'e nuwe rekenmodel, soortgelyk aan die universele Turing-masjien. Deutsch het ook 
die benadering deur kwantumlogiese stroombane met behulp van die \emph{Deutsch-hek} 
(kwantumeweknie van die Toffoli-hek) gevestig.

\section{\textsc{Uit die Sowjet-kubernetika}}

Die bydrae van die Sowjetunie op die gebied van die teoretiese rekenaarwetenskap en verwante 
gebiede was groot. Min wetenskaplikes sal nie met die meeste van die name 
L.V. Kantorovitch\footnote{
Gebruik van rekenaars in optimering, veral in ekonomiese beplanning.}, 
A.P. Ershov\footnote{
Teoretiese en stelselprogrammering, saamsteller-ontwerp, in 1974 benoem tot 
\emph{Distinguished Fellow of the British Computer Society}.
}, P.S. Novikov\footnote{
Bewys in 1952 dat die woordprobleem vir groepe onoplosbaar is, werk in die wiskundige 
logika en teorie van algoritmes.}, S.A. Lebedev\footnote{
Inwerkingstelling van die eerste rekenaar met gestoorde program in die Sowjetunie en ook op die 
Europese vasteland in Ki\"ef in 1951 \cite{MathMachine}.}, A.N. Kolmogorov\footnote{
Grondlegging van waarskynlikheidsteorie, konsipiering van die Kolmogorov-kompleksiteitsmaat, onder vele andere.}, 
L.A. Levin\footnote{
Toevalligheid, algoritmiese inligtingsteorie, berekenbaarheid.} of 
Yu.V. Matiyasevich\footnote{Oplos van Hilbert se Tiende Probleem.} vertroud wees nie. Dit 
was veral onder die vaandel van die \emph{kubernetika} wat hierdie wetenskaplikes gewerk het.

Kubernetika, waarvan die term deur Norbert Wiener in 1948 bekend gemaak is in sy invloedryke 
werk \emph{Cybernetics, or Control and Communication in the Animal and the Machine}, het in die 
Weste in die dekades na die verskyning van di\'e werk versplinter in die dissiplines 
van operasionele navorsing,  stelselteorie, beheerteorie en inligtingsteorie. Ten spyte van 
'n minder gunstige begin as in die VSA, het die kubernetika in die USSR 'n groot impak op 
die wetenskaplike gemeenskap gehad. 
In die Sowjetunie is die kubernetika in die tydperk onmiddellik na die Tweede W\^ereldoorlog  
as 'n instrument van die VSA se imperialistiese ideologie en as 'n pseudowetenskap 
afgemaak \cite{MathMachine}. Derhalwe moes Sowjet-wetenskaplikes besonder versigtig omgaan met 
terminologie, terwyl hulle hard gewerk het aan rekenaarstelsels vir milit\^ere 
doeleindes. Die vertaling van die bogenoemde bekende boek van Norbert Wiener is 
byvoorbeeld met tien jaar vertraag \cite{MathMachine} weens die kwansuis spekulatiewe en 
filosofiese aard daarvan op 'n stadium toe die owerhede reeds aktief die vertaling van Westerse 
vakliteratuur na Russies bevorder het. Die sirkulasie van enkele eksemplare van di\'e 
boek het egter die regte klimaat geskep vir 'n oplewing in die kubernetika en 
teoretiese rekenaarwetenskap na Stalin se dood in 1953. Di\'e ideologiese ommekeer was so 
groot dat die kubernetika in 1961 in die Kommunistiese 
Party se program opgeneem is as 'n sleutelwetenskap vir die skepping van die materi\"ele 
en tegniese grondslae van kommunisme \cite{Scandals}. 

Die \emph{Great Soviet Encyclopedia} \cite{GSE} definieer kubernetika kort- en breedweg, in 
die Engelse vertaling van sy derde uitgawe, as 
"`Cybernetics, the science of control, communications and data processing.''  
Hoewel die USSR reeds teen die 
middel 1960's 'n agterstand van minstens vyf jaar teenoor die VSA gehad het 
 op die gebied van rekenaartegnologie en al verder en 
verder veld verloor het \cite{Klimenko}, het 
die 1960's en veral die vroe\"e 1970's 'n bloeitydperk vir die kubernetika en 
rekenaarwetenskap in die Sowjetunie geword. Binne die\,---\,skielik polities korrekte\,---\,kubernetika 
was daar ruimte vir die wiskundige logika en teorie van berekening om te ontwikkel n\'a dekades 
van ontmoediging. 
Teen die 1980's het kubernetika ook in die 
Sowjetunie as wetenskap doodgeloop, deels onder Westerse invloed en deels weens 'n te 
noue assosiasie met die staatsideologie \cite{Vucinich}. 
Di\'e negatiewe verwikkeling in die studie van kubernetika in die USSR is moontlik deels 
te blameer daarvoor dat Yuri\footnote{
Daar bestaan 'n Babelse verwarring om die transliterasie van name uit die Russies. Volgens die 
standaard ISO 9 (1995) skryf mens vir die algemene naam {\cyr Yurii0} in die Latynse alfabet 
\emph{\^Urij}, maar d\'it word selde of nooit so aangetref. Die outeur volg 
in die hoofteks hier die gewoonte om di\'e naam \emph{Yuri} (soos -- Gagarin) te skryf, wat ook 
gebruiklik is onder Russe self wat s\'o heet. Die Afrikaanse fonetiese transkripsie word 
eerder heeltemaal vermy, maar in die verwysings kom die Hongaarse transkripsie voor soos 
gebruik in di\'e spesifieke bron.}  
Manin se opmerkings in sy boek \emph{Berekenbaar en Onberekenbaar} \cite{Manin1980} 
(sover di\'e outeur kon vasstel) nie vertaal en wyer gelees is nie. Di\'e boek 
bevat \cite{ZB} 'n uittreksel van die Russiese uitgawe van die bekende handboek van Manin, 
\emph{A~Course in Mathematical Logic} (reeds in 1977 in Engels vertaal).
Manin skryf in 1980 \cite{Manin1980} ($\C$ en $c$ staan vir die versameling komplekse getalle), 

\smallskip

\begin{Parallel}{8.75cm}{8.5cm}
\small
\ParallelLText{\noindent 
{Dit is moontlik dat vir die beter begrip van sulke verskynsels 'n wiskundige teorie 
van kwantumoutomate ontbreek.  Die wiskundige model van sulke objekte moet heel 
vreemde eienskappe vertoon in vergelyking met di\'e van deterministiese prosesse. 
Hiervoor is die rede dat die kapasiteit van die kwantumruimte dramaties groter 
is: indien daar in die klassieke geval $N$ diskrete toestande is, dan is in die 
kwantumteorie (volgens die superposisiebeginsel) die toestandruimte in $\C^N$. 
In klassieke stelsels het die samevoeging van stelsels met \mbox{$N_1$} en \mbox{$N_2$} toestande 
onderskeidelik bloot die produk aantal toestande, maar in die kwantumgeval is dit 
\mbox{$\C^{N_1N_2}$}.
}
}
\ParallelRText{\noindent 
Lehets\'eges, hogy az ilyen jelens\'egek jobb meg\'ert\'es\'ehez a
kvantumautomat\'ak matematikai elm\'elete hi\'anyzik. As ilyen objektumok 
matematikai modellt mutathatn\'anak az eg\'eszen szokatlan tulajdons\'agokkal 
rendelkez\H o determinisztikus folyamatokra. Ennek egyik oka az, hogy a kvantumt\'er 
t\'erfogata l\'enyegesen nagyobb a klasszikusok\'en\'al: ott, ahol a klasszikusban \mbox{$N$} 
diszkr\'et \'allapot van, a szuperpoz\'ici\'ojukat megenged\H o kvantumelm\'eletben $c^N$ 
elemi t\'erfogat van. A klasszikus rendszerek egyes\'it\'esikor az \mbox{$N_1$} 
\'es \mbox{$N_2$} 
\'allapotsz\'amok \"osszeszorz\'odnak, a kvantumos v\'altozatban \mbox{$c^{N_1N_2}$} 
ad\'odik.
}
\ParallelPar
\ParallelLText{
{Hierdie growwe berekeninge toon dat di\'e stelsels met kwantumgedrag potensieel ver 
meer ingewikkeld is as die klassieke weergawe. Byvoorbeeld: omdat die stelsels geen unieke  
regte dekomposisie het nie, kan die toestand van die kwantumoutomaat op vele maniere 
waargeneem word as heeltemaal ver\-skillende toestande van virtuele klassieke outomate. 
Aan die einde van [17] byvoorbeeld kan 'n regtig leersame berekening gevind word: 
vir die kwantummeganiese beskrywing van die metaanmolekule moet waardes op 'n tralie in 
$10^{42}$ punte uitgereken word. Indien ons aanneem dat in elke punt in totaal 10 element\^ere 
bewerkings uitgevoer moet word en veronderstel dat elke bewerking  
teen werklik lae temperatuur plaasvind ($T=3\cdot10^{-3}~\mbox{K}$), dan moet ons soveel 
energie verbruik as wat op Aarde in ongeveer een eeu opgewek word.
}
}
\ParallelRText{
Ezek a durva sz\'am\'it\'asok megmutatj\'ak, hogy a rendszerek kvantumos viselked\'ese 
potenci\'alisan sokkal bonyolultabb a klasszikus ut\'anzatok\'en\'al. P\'eld\'aul, amiat 
hogy a rendszernek nincs elemekre val\'o egy\'ertelm\H u felbont\'asa, a kvantumautomata 
\'allapota sokf\'elek\'eppen tekinthet\H o teljesen k\"ul\"onb\"oz\H o virtu\'alis 
klasszikus automat\'ak \'allapotainak. Pl. a [17] v\'eg\'en tal\'alhat\'o egy igen 
tanuls\'agos sz\'amit\'as: A met\'anmolekula kvantumechanikai sz\'ambav\'etel\'ehez 
h\'al\'o m\'odszerrel $10^{42}$ pontban kell sz\'am\'it\'ast v\'egezni. Ha \'ugy 
vessz\"uk, hogy minden pontban \"osszesen 10 elemi m\H uveletet kell elv\'egezni, \'es 
feltessz\"uk, hogy minden sz\'am\'it\'as igen kis h\H om\'ers\'ekleten megy v\'egbe 
($T=3\cdot10^{-3}~\mbox{K}$), akkor ennyi energi\'at kell felhaszn\'alnunk, amennyit a 
F\"old\"on hozz\'avet\H oleg egy \'evsz\'azad alatt \'all\'itanak el\H o.
}
\ParallelPar
\ParallelLText{
{Tydens 'n dergelike verwesentlikingsprogram sal die eerste struikelblok wees om die 
juiste balans te vind vir die wiskundige en fisiese beginsels. Kwantumoutomate sou  
abstrak wees\,---\,die wiskundige modelle moet slegs voldoen aan die algemene beginsels 
van die kwantumteorie, sonder fisiese implementasie. Nou is die evolusie van die model 'n 
unit\^ere rotasie in eindigdimensionele Hilbert-ruimte en die virtuele dekomposisie in 
deelstelsels stem ooreen met die ruimte se tensorproduk-dekomposisie. Iewers moet steeds in 
hierdie prentjie die plek gevind word van interaksies, wat tradisioneel met Hermitiaanse 
operatore en waarskynlikhede beskryf word. 
(Outeur se vertaling uit die Hongaarse vertaling van die oorspronklike.)}
}
\ParallelRText{
Egy ilyen program megval\'os\'it\'asa sor\'an az els\H o neh\'ezs\'eg a matematikai 
\'es fizikai elvek megfelel\H o egyens\'uly\'anak megtal\'al\'asa. A kvantumautomat\'anak 
abszraktnak kell lennie, a matematikai modellnek csup\'an az \'altal\'anos kvantumelm\'eleti 
elveket kell kiel\'eg\'itenie, a fizikai realiz\'aci\'o meghat\'aroz\'asa n\'elk\"ul. Ekkor  
az evol\'uci\'o modellje unit\'er forgat\'as a v\'eges dimenzi\'os Hilbert-t\'erben, a 
r\'eszrendszerekre val\'o virtu\'alis felbont\'as pedig a t\'er tenzorszorzatt\'a val\'o 
felbont\'as\'anak felel meg. Valahol meg kell tal\'alni eben a k\'epben a k\"olcs\"onhat\'asok 
hely\'et, amiket tradicion\'aliian Hermite-oper\'atorokkal \'es -val\'osz\'in\H us\'eggel 
\'irnak le.
}
\end{Parallel}

\smallskip
\noindent
Manin wys duidelik hier op die belangrikheid van die superposisiebeginsel vir 
kwantumberekening, soos Feynman twee jaar later ook sou doen. 
In die Russiese vakliteratuur word daar algemeen verwys na hierdie opmerkings in 1980 
van Manin as die vroegste spekulasie op skrif oor die onderwerp van 
kwantumberekening. In \cite{Guts}, byvoorbeeld, vermeld A.K. Guts:

\smallskip

\begin{Parallel}{8.75cm}{8.5cm}
\small 
\ParallelLText{ \noindent
{In die jaar 1980 het Yu.I. Manin gewys op die onvermydelike behoefte aan 
'n teorie van kwantumrekentoestelle. (Outeur se vertaling)
}
}
\ParallelRText{\noindent 
\cyr 
V 1980 g. Yu.I. Manin ukazal na neobkhodimostp1 razrabotki teorii kvantovykh vychislitelp1nykh 
ustroi0stv.
} 
\end{Parallel}

\smallskip
\noindent
N\'og meer duidelik skryf Yu.G. Neizvestniy \cite{Nei}, byvoorbeeld:
\smallskip 
\begin{Parallel}{8.75cm}{8.5cm}
\small 
\ParallelLText{\noindent 
{Die idee van kwantumberekening is die eerste keer geopper deur Yu.I. Manin in die jaar 1980, 
maar hierdie probleem het eers aktief onder bespreking gekom na die verskyning van 
'n artikel deur die Amerikaanse teoretiese fisikus R. Feynman in 1982. In hierdie werke is die 
gebruik van die toestande van 'n kwantumstelsel vir berekeningsoperasies voorgestel. 
(Outeur se vertaling)
}
}
\ParallelRText{\noindent 
\cyr
Ideya kvantovykh vychislenii0 vpervye byla vy-kazana Yu. I. Maninym v 1980 godu, no 
aktivno e1ta problema stala obsuzhdatp1sya posle poyavleniya v 1982 godu statp1i amerikanskogo 
fizika-teoretika R. Fei0nmana. V e1tikh rabotakh bylo predlozheno ispolp1zobatp1 
dlya vychislennii0 operatsii s sostoyaniyami kvantovoi0 sistemy.
} 
\end{Parallel}

\smallskip
\noindent
Di\'e erkenning aan Manin vir sy idee word egter ongelukkig nog selde in die Westerse vakliteratuur 
aangetref, met enige uitsonderings soos Manin se eie uitstekende oorsig van kwantumalgoritmes 
\cite{Manin1999} of Knill e.a. se inleidende teks \cite{KnillEA}. 'n Gedeeltelike 
vertaling in Engels van Manin se opmerkings kan gelees word op 'n kursustuisblad\footnote{
\url{http://iaks-www.ira.uka.de/home/grassl/Academia/QECC/zitate.html}} 
van die Universiteit van Karlsruhe. Verder is daar enkele verwysings in vertalings 
uit Russies, byvoorbeeld die leerboek \cite{KitaevEA} van Kitaev e.a.

\section{\textsc{Slotopmerkings en gevolgtrekking}}

In die vakliteratuur was Paul Benioff die eerste om 'n gedetailleerde model vir 'n 
rekenmasjien gebaseer op die beginsels van die kwantummeganika te beskryf. Benioff 
behoort as di\'e pionier van kwantumberekening beskou te word op grond van sy bydraes. 
Yuri Manin het onafhanklik en in dieselfde jaar in \cite{Manin1980} gewys op die 
moontlikheid van sulke masjiene \'en op die krag wat opgesluit l\^e in die superposisiebeginsel. 
Weens die redes hierbo genoem was die impak van Manin se stellings 
buite die Sowjetunie weining of min. 
'n Bietjie later het Richard Feynman ook op die krag van die superposisiebeginsel 
begin wys en met baie groot sukses di\'e idees aan die wetenskaplike publiek verkondig. 
Nieteenstaande Feynman se impak behoort die wetenskaplike publiek die vroe\"er 
gepubliseerde werk van Benioff en  Manin as die ware begin van kwantumberekening 
in die vakliteratuur te erken.

Die Inligtingsparadoks illustreer op betreklik dramatiese wyse die wisselwerking 
tussen inligtingsteorie en die grondslae van die fisika. 
Om formeel te dink oor berekening kan ons insae gee in die fisika (soos in die geval 
van Bennett se oplossing van die skynparadoks van Maxwell se Duiweltjie) en sulke 
denke kan ook gebruik word om fisiese teorie\"e te yk, soos in die geval van 
die Inligtingsparadoks. Hoewel kwantumrekenaars niks kan doen 
wat nie op 'n konvensionele rekenaar of Turing-masjien stadig nageboots kan word nie \cite{Davis}, 
is die \emph{bewerkingspoed} van kwantumrekenaars van so 'n aard dat alle toegepaste 
wetenskaplikes hulle konsep van wat in praktyk uitrekenbaar is, sal moet herdink 
indien\,---\,of wannneer\,---\,die tegnologie vir die produksie van kwantumrekentoestelle op 
redelike skaal ontwikkel is. Teoretiese rekenaarwetenskaplikes moet ook 'n duidelike 
definisie van \emph{berekening} h\^e en hoewel kwantumrekenaars niks hieraan sal 
verander nie, kan 'n omwenteling in die sienings van die fisika 'n groot 
uitwerking op die rekenaarwetenskap h\^e, al is dit aanvanklik net konseptueel. Veral 
die snaarteorie het baie duidelike aanknopingspunte met inligtingsteorie en 'n 
goeie aanslag op die Inligtingsparadoks deur snaarteoretici, behoort van groot belang 
vir ons te wees.


\section*{Kort biografie van die outeur}

Petrus Potgieter (jaargang 1968) verwerf in 1992 die graad M.A. aan Kent State University in 
die VSA en in 1996 die graad Ph.D. in wiskunde aan die Universiteit van Pretoria. Hy doseer twee 
jaar aan die Universiteit van Stellenbosch en is sedert 1997 verbonde aan die Departement 
Kwantitatiewe Bestuur by die Universiteit van 
Suid-Afrika, tans as medeprofessor. Sy hoofbelangstellings in die akademie is die 
grondslae van berekening en die wiskunde van finansies. Die temas wat hierdie studievelde 
verbind is eerstens die verband van beide met die waarskynlikheidsteorie en 
tweedens die siening dat albei hulself besig hou met die koppelvlak tussen 'n agent 
(rekentoestel \'of belegger) en 'n stelsel (menslike waarnemers \'of  die ekonomie) deur 
middel van goed gedefinieerde seine (invoer/uitvoer \'of  pryse, respektiewelik).

\end{document}